\documentclass[aps,twocolumn,nofootinbib]{revtex4}
\usepackage{graphicx}% Include figure files
\usepackage{dcolumn}% Align table columns on decimal point
\usepackage{bm}% bold math
\usepackage{amsmath}
\usepackage{amssymb} 

\usepackage{float}

\begin{document}

\title{Quantum Vacuum Experiments Using High Intensity Lasers}

\author{Mattias Marklund}
\email{mattias.marklund@physics.umu.se}
\affiliation{Department of Physics, Ume{\aa} University, SE--901 87 Ume{\aa}, Sweden}

\author{Joakim Lundin}
\affiliation{Department of Physics, Ume{\aa} University, SE--901 87 Ume{\aa}, Sweden}

\date{November 17, 2008}

\begin{abstract}
  The quantum vacuum constitutes a fascinating medium of study, in particular since near-future laser facilities will be able to probe the nonlinear nature of this vacuum. There has been a large number of proposed tests of the low-energy, high intensity regime of quantum electrodynamics (QED) where the nonlinear aspects of the electromagnetic vacuum come into play, and we will here give a short description of some of these. Such studies can shed light, not only on the validity of QED, but also on certain aspects of nonperturbative effects, and thus also give insights for quantum field theories in general.
\end{abstract}

  \pacs{
}

\maketitle

\section{Introduction}

The event of high-power lasers has dramatically changed the way we do science. As a tool for basic and applied research, lasers are surpassed by few in its usefulness and breath. Spectroscopic techniques based on laser systems gives us high-precision measurements, lasers and optical fibers gives a means of rapid long range communication, and the prospects of using laser accelerated protons for hadron theory is within reach in the near future, to name but a few examples. For fundamental physics, the coherent nature of lasers in conjunction with new compression techniques gives us an unsurpassed opportunity to investigate the ultra-high intensity realm of physical theory. In particular, quantum electrodynamics (QED) can soon be probed in its non-perturbative regime, giving important information about the theory where theory is still struggling. Furthermore, such tests of QED can also be useful for understanding the inherent non-perturbative nature of other quantum field theories. But perhaps the most exciting perspective is the possibility to find new physics in such high-intensity environment, as well as constraining modifications of current standard models \cite{Gies} (see Fig. 1).

%%%%% FIG %%%%%
\begin{figure}
  \includegraphics[width=.9\columnwidth]{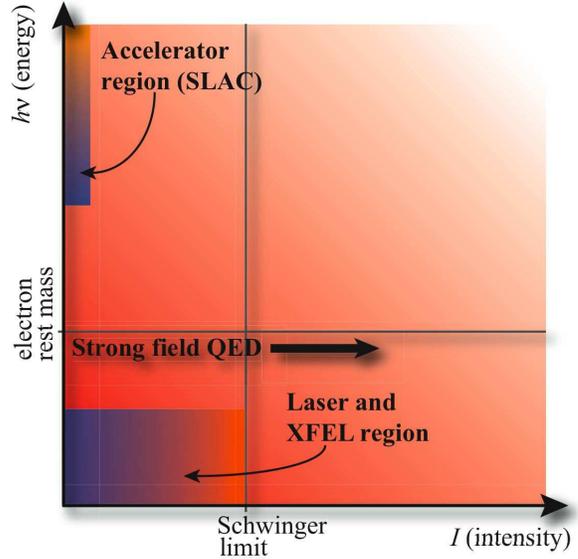}
  \caption{A picture of the different regimes covered by accelerators and lasers (not to scale).}
\end{figure}
%%%%%%%%%%%%

In this paper we will look into the experimental prospects of doing laser based experiments on the quantum vacuum. In particular, we will discuss a few of the many suggestions in the literature. Thus, the presentation in this paper is in no way conclusive, and does not aim at covering all the important aspects of this growing field of research. However, we hope that the material at hand will give a flavor of the current status of quantum vacuum investigations. 

In closing of this introduction, the following quotation might seem prudent computational methods \cite{Mattuck}
\begin{quote}
In eighteen-century Newtonian mechanics, the three-body problem was insoluble. With the birth of general relativity around 1910 and quantum electrodynamics in 1930, the two- and one-body problems became insoluble. And within modern quantum field theory, the problem of zero bodies (vacuum) is insoluble.
\end{quote}	

\section{The quantum vacuum}
Quantum field theory (QFT) allows virtual particle pairs to be spontaneously created in vacuum provided that they annihilate each other within a sufficiently short time governed by the Heisenberg uncertainty principle. Since fluctuations of virtual particles will always be present, this leads to a redefinition of our concept of what constitutes a vacuum. A vacuum is the lowest energy state of a system, which usually implies a volume in absence of real particles. The vacuum fluctuations do in general not give note of themselves unless the vacuum is disturbed in some way. The disturbance can be in the form of an external electromagnetic field or sometimes in the form of simple boundary conditions. Below we briefly outline a few effects arising from the non-classical properties of the quantum vacuum.\\
\\
\textbf{The Casimir effect}\\
Virtual photons must obey the same boundary conditions as classical fields. So, if two parallel perfectly conducting plates are placed close to each other in vacuum, only standing wave modes are allowed to exist in between the plates, whereas any mode can exist outside where the boundaries are at infinity. The vacuum energy density in between the two plates will then be lower than outside, and there will be a net attractive force between the plates \cite{Casimir1,Casimir2} (see Fig.\ 2). 
This effect has been successfully verified in experiment by e.g.\ Refs.\ \cite{Sukenik1993,Mostepanenko1997,Lamoreaux1998,Bordag2001,Bressi2002,Harber2005}, see also references therein.\\
%%%%% FIG %%%%%
\begin{figure}
  \includegraphics[width=.9\columnwidth]{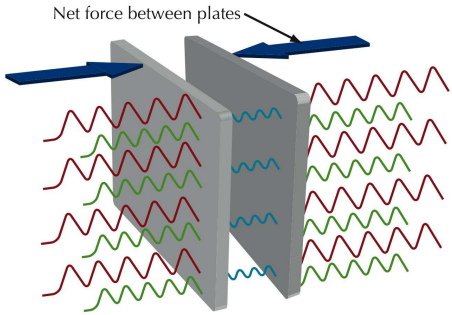}
  \caption{A heuristic representation of the Casimir effect.}
\end{figure}
%%%%%%%%%%%%
\\
\textbf{Pair creation}\\
	Suppose that the vacuum is disturbed by a strong external electric field. A virtual electron-positron pair may then become real if the work exercised by the electric field on an electron over the distance of approximately a Compton wavelength is of the order of the electron rest mass (see Fig.\ 3). Pair production has been observed in experiment where high frequency photons interacted with an intense electromagnetic field, \cite{Burke} (see Fig.\ 4). Pair creation by focused laser beams has also been discussed in e.g.\ Refs.\ \cite{Alkofer2001,Narozhny2004}.\\
%%%%% FIG %%%%%
\begin{figure}
  \includegraphics[width=.9\columnwidth]{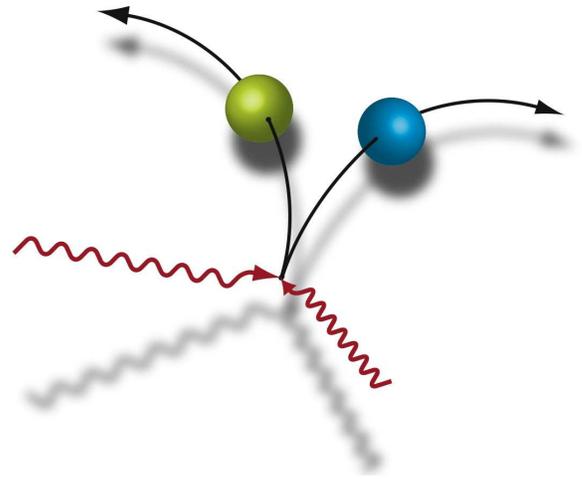}
  \caption{Pair production through photon interaction.}
\end{figure}
%%%%%%%%%%%%
%%%%% FIG %%%%%
\begin{figure}
  \includegraphics[width=.9\columnwidth]{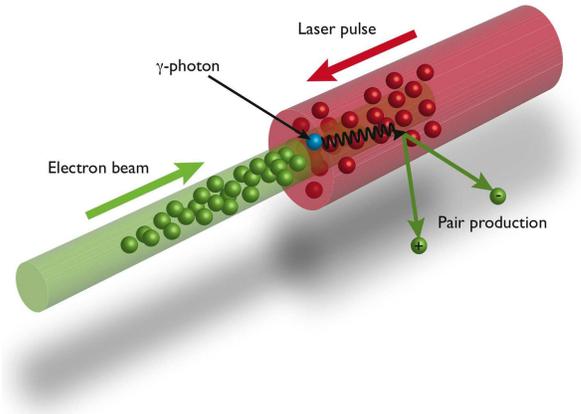}
  \caption{A schematic picture of the experiment presented in Ref. \cite{Burke}.}
\end{figure}
%%%%%%%%%%%%
\\
\textbf{Photon-photon scattering}\\
	Photons are indifferent to each other within classical electrodynamics in the absence of a material medium. Two laser beams colliding in vacuum would simply not affect each other. However, photons can interact with the virtual electron-positron pairs of the quantum vacuum. In this way, an energy and momentum exchange between the photons can be mediated by virtual particle pairs of quantum vacuum fluctuations. Thus, two or more photons may scatter off each other (see Fig.\ 5).\\
%%%%% FIG %%%%%
\begin{figure}
  \includegraphics[width=.9\columnwidth]{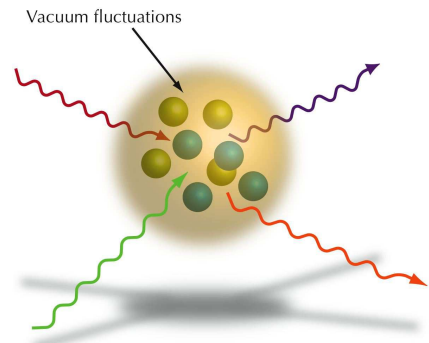}
  \caption{The process of elastic scattering among photons through the quantum vacuum.}
\end{figure}
%%%%%%%%%%%%
	\\
\textbf{Vacuum birefringence}\\
	A strong external magnetic field will affect the virtual electron-positron fluctuations of the quantum vacuum. As a result, the phase velocity of photons propagating in the ordinary mode and the extraordinary mode will be different \cite{Heyl1997,Heinzl2008}.\\
\\	
\textbf{Photon acceleration in vacuum}\\
A test photon propagating along a plasma oscillation will experience different plasma densities depending on the position of the photon. The density perturbations can affect the group velocity of the test photon, and for this reason the process is called photon acceleration. In the same way, a test photon propagating in an intense modulated background of radiation, e.g.\ in the form of a nearly unidirectional photon beam, will experience a frequency shift. Here the plasma density perturbations are replaced by a trail of density perturbations of virtual electron-positron pairs induced by the intense background radiation itself \cite{Mendonca, Mendonca2006}.\\
\\	
\textbf{Self-lensing effects}\\
	A strong pulse of light may substantially affect the properties of the vacuum in which it propagates, thus effecting its own evolution. However, as a consequence of Lorentz invariance the lowest order non-linear terms describing this self-interaction vanish for parallel propagation. So, the vacuum needs to be modified by an external electromagnetic field or with a wave guide for this effect to become important. This self-interaction can have a self-lensing effect on the pulse if the vacuum is properly modified, and this can lead to effects such as photon splitting in the presence of an external magnetic field \cite{Adler1971}, the formation of light bullets\footnote{No stable soliton solution to the ordinary (i.e. cubic) nonlinear Schr\"odinger equation in dimensions larger than one have been found (see e.g.\ Ref.\ \cite{Desaix1991}), but neither has a proof stating that this cannot be found. The question whether higher dimensional stable photon bullets can exist or not thus still remains open (however, see Ref.\ \cite{Belic2008}).} in suitable waveguides \cite{Brodin2003} or properly modulated background fields \cite{Soljacic}, or pulse collapse in an intense gas of photons \cite{Marklund2003}-\cite{Rozanov1998}.\\
\\
Most of the quantum vacuum effects listed above do not become important unless the vacuum is disturbed by a strong electric and/or magnetic field (the Casimir effect is an exception). The static electric field strength threshold above which quantum vacuum effects no longer can be neglected is often set to be the Sauter-Schwinger limit, where $E_{\textrm{crit}}\approx 10^{16}\ \mathrm{Vcm}^{-1}$. This field strength immediately translates into a critical intensity, $10^{29}\ \mathrm{Wcm}^{-2}$. Above this intensity, the vacuum is no longer stable and we can expect significant electron-positron pair production. This threshold may be relaxed if the field considered is not static but time-varying. For the case of a modulated laser pulse, it is expected that the threshold intensity will be reduced by 1-2 orders of magnitude \cite{Narozhny2004}. Currently, lasers can reach intensities of around $10^{21}-10^{22}\ \mathrm{Wcm}^{-2}$ \cite{Bahk2004, Mourou2006}, but the laser power is expected to continue to increase for some time \cite{Mourou2006,Mourou1998}. For instance, there are prospects of laser systems, e.g.\ the Extreme Light Infrastructure (ELI) \cite{ELI} and the High Power laser Energy Research system (HiPER) \cite{HiPER}, that offers the potential of reaching intensities exceeding $10^{25}\ \mathrm{Wcm}^{-2}$. Such intensities would truly challenge the vacuum critical field, and perhaps open up for direct studies of the quantum vacuum.

\section{Why lasers?}
QED is a well tested theory, in particular in the high energy low intensity regime of particle accelerators (CERN, SLAC etc.). The theory, however, is not very well tested in the soft photon high intensity regime. This has mainly been due to the limit of available high field strengths. High-$Z$ atoms offer high electric field strengths and have been used to detect Delbr\"{u}ck scattering \cite{Jarlskog1973} and photon splitting \cite{Akhmadaliev2002}. Many QED processes, however, still remain untested in this high field regime. Photon-photon scattering is one such example, and it can only become experimentally accessible in this high intensity regime because of the small scattering cross-section (of the order of $10^{-65}\ \textrm{cm}^2$ in the optical regime). With the rapidly growing powers of present day laser systems, intensity regimes where processes like photon-photon scattering and electron-positron pair creation may be of importance is expected to become within reach in a near future. 

Quantum vacuum experiments using external fields have also been proposed as a way of probing new physics \cite{Gies,Gies2008}. These kind of experiments are particularly suitable for searching for new low energy weakly interacting degrees of freedom, e.g.\ light pseudo scalar particles such as axions\footnote{Axions are bosons which were introduced in order to explain the absence of CP symmetry breaking in QCD \cite{Peccei-Quinn, Weinberg,Wilczek}, and the axion is still to be detected.}. It is possible that in the future, quantum vacuum experiments can be used to explore parameter ranges of particle physics beyond the Standard Model. Laser driven experiments are thus an important complement to accelerator driven experiments.\\

\section{Photon--photon scattering}

\subsection{Experimental four-wave mixing}

Direct observation of elastic photon--photon scattering among real photons would be an important benchmark test of laser based QED experiments. Deviations for the expected scattering rate would indicate new physics in the low-energy regime. Throughout the last decades, several suggestions on how to detect elastic photon--photon scattering, using laser assisted schemes, have been made. For instance, Refs.\ \cite{Ferrando2007, Tommasini2008} suggest that non-linear interactions between colliding laser pulses in vacuum will lead to a measurable phase shift at exawatt power regiems. Moreover, crossing electromagnetic waves can interact and yield new modes of different frequencies. One of the more prominent modes in such a mechanism is given by the four-wave interaction mediated mode satisfying resonance condition between the frequencies and wavevectors (i.e. photon energy and momentum conservation) \cite{Rozanov93}. It is therefore not a surprise, given the evolution of laser powers and frequencies, that the search for photon--photon scattering using resonant four-wave interactions has caught the attention of researchers in this area. This approach has also come furthest in the experimental attempts to detect elastic scattering among photons \cite{Bernard,Moulin-Bernard,Bernard2,Bernard-etal,Bernard3}.  

Using the resonance conditions $\omega_4 = \omega_1 + \omega_2 - \omega_3$ and $\mathbf{k}_4 = \mathbf{k}_1 + \mathbf{k}_2 - \mathbf{k}_3$, between the vacuum generated photons and the laser pump sources, respectively, one may derive a set of wave interaction equations for slowly varying amplitudes $a_i, \, i = 1, \ldots, 4$, of the form
\cite{Marklund-Shukla}
\begin{equation}
  \frac{da_i}{dt} = Ca_ja_ka^{\ast}_l ,
\end{equation}
given any type of media through which the waves may interact. Here the coupling constants $C$ depend on the interaction in question, as well as on the physical parameters of the system around which the waves are modulated.

The coupling constants may be interpreted in terms of the nonlinear susceptibility of the vacuum. Moulin \& Bernard \cite{Moulin-Bernard} considered the interaction of three crossing waves, characterized by their respective electric field vectors $\mathbf{E}_i$, producing a fourth wave $E_4$. Starting from Maxwell's equations with the usual weak field limit Heisenberg--Euler third order nonlinear corrections, they derive the nonlinear Schr\"odinger equation (see also Ref.\ \cite{Marklund2003} for similar results in a different setting)
\begin{equation}
 i\left( \frac{\partial}{\partial t} + c\frac{\partial}{\partial z} 
 \right)E_4 + \frac{c^2}{2\omega_4}\nabla^2_{\perp}E_4 = 
 -\frac{\omega_4}{2}\chi^{(3)} E_1E_2E_3^{\ast}
\end{equation}
for the driven wave amplitude $E_4$, where the overall harmonic time dependence $\exp(-i\omega t)$ has been factored out. Here $\chi^{(3)}$ is the third order nonlinear susceptibility given by
\begin{equation}
  \chi^{(3)} %= \frac{2\hbar e^4}{360\pi^2m_e^4c^7\epsilon_0}K
  = \frac{\alpha}{45\pi}\frac{K}{E_{\text{crit}}^2}
  \approx 3\times 10^{-41}\times K \, \mathrm{m^2/V^2} ,
\end{equation}
where $K$ is a dimensionless form factor of order unity. The value of $K$ depends on the polarization and propagation directions of the pump modes, and reaches a maximum of $K = 14$ for degenerate four-wave mixing \cite{Moulin-Bernard}. Refs.\ \cite{Bernard3} and \cite{Bernard-etal} presented experiments on four-wave mixing in vacuum, improving previous attempts by nine orders of magnitude, although no direct detection of photon--photon scattering was achieved. 

\begin{figure}[ht]
\includegraphics[width=0.9\columnwidth]{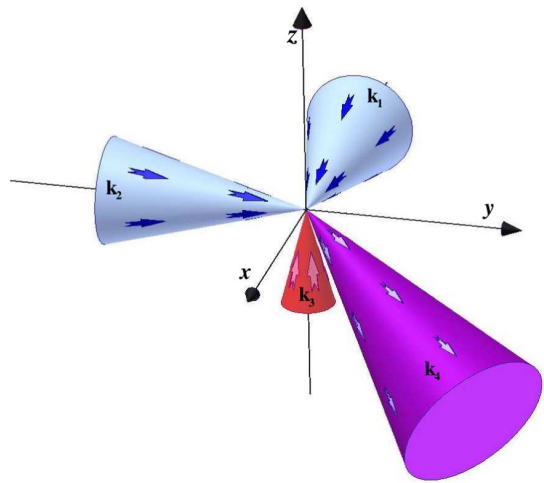}
\caption{Configuration of the incoming laser beams (represented by the wave vectors $\mathbf{k}_1, \mathbf{k}_2$ and $\mathbf{k}_3$) and the direction of the scattered wave (with wave vector $\mathbf{k}_4$).}
\label{fig:3d}
\end{figure}

\begin{figure}[H]
\begin{center}
\begin{minipage}{0.49\linewidth}
\includegraphics[width=1\textwidth]{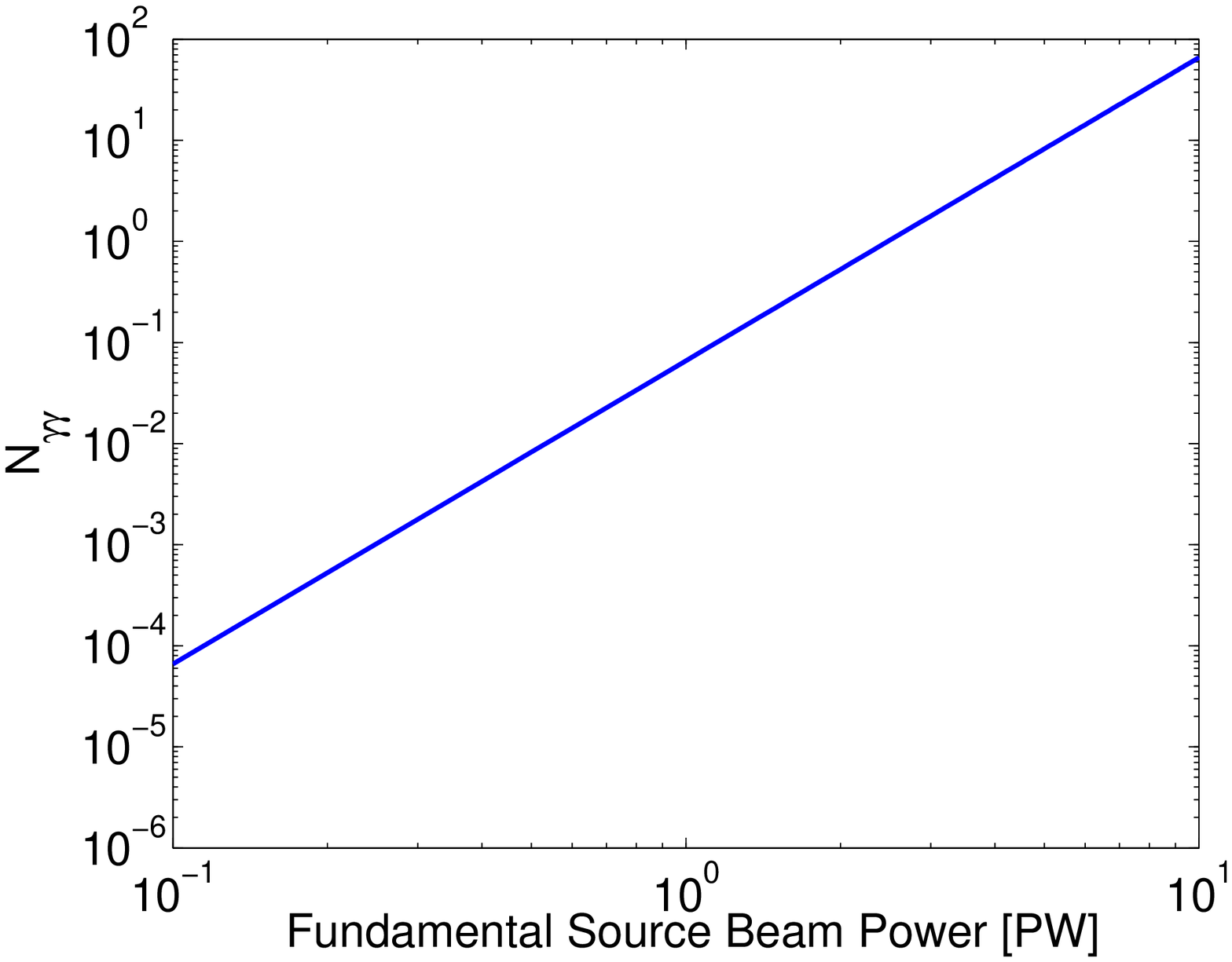}
\begin{footnotesize}
\begin{center}
\ \\
\textbf{(a)}
\end{center}
\end{footnotesize}
\end{minipage}
\begin{minipage}{0.49\linewidth}
\includegraphics[width=1.2\textwidth]{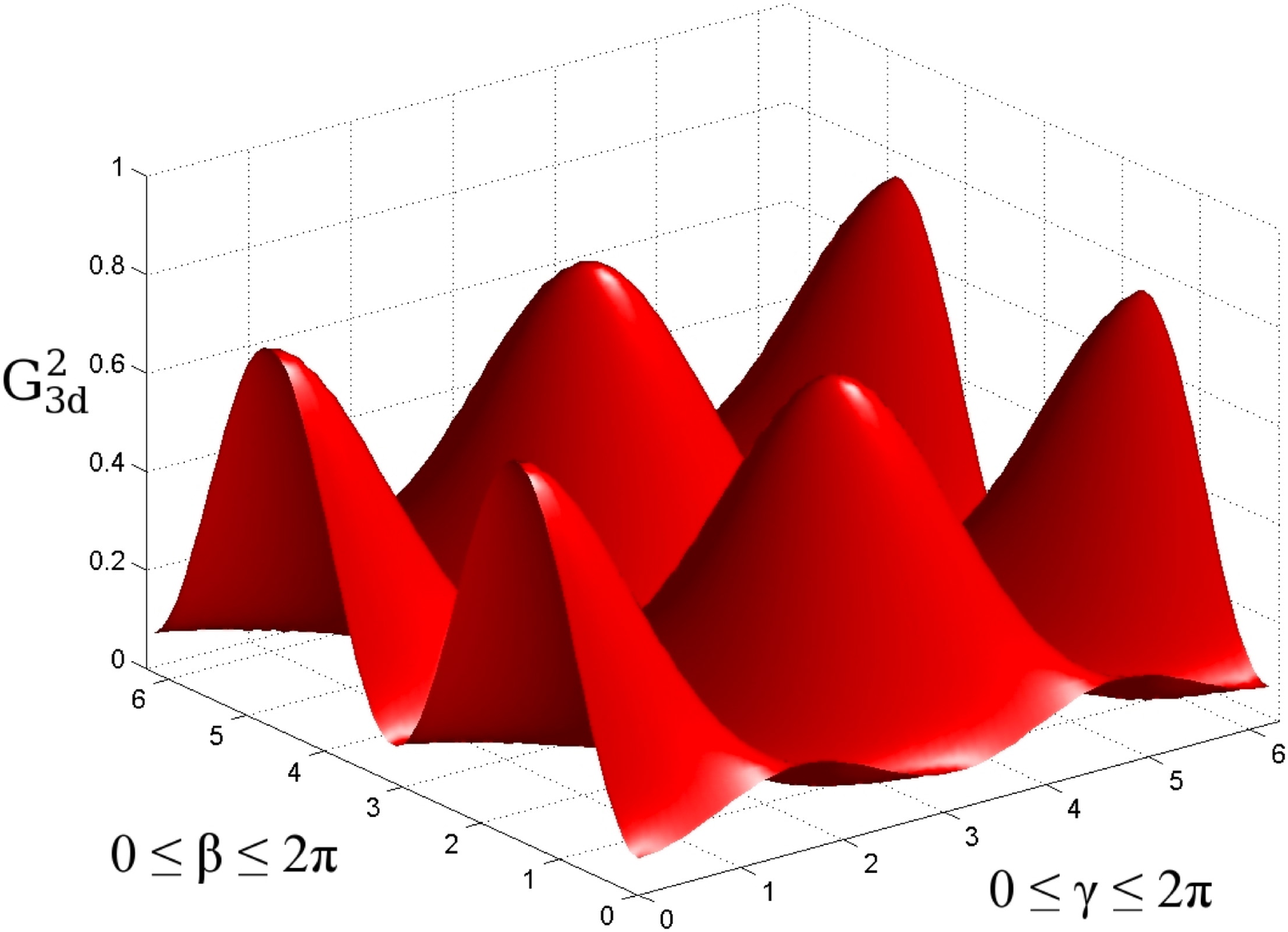}
\begin{footnotesize}
\begin{center}
\textbf{(b)}
\end{center}
\end{footnotesize}
\end{minipage}
\caption{$(a)$ shows the expected number of generated photons predicted by (\ref{eq:number}) when increasing the laser power while keeping the beam width constant at $b=1.6\ \mathrm{\mu m}$. The beam power is that of a single beam that is split, frequency doubled (with 60\% energy loss) and then split again in order to obtain the configuration in Fig.\ (\ref{fig:3d}). Numbers are calculated assuming an optimum choice of polarizations. $(b)$ illustrates the polarization dependence of the signal when the polarization of one beam is kept fixed at an optimal value, while the polarization of the other two beams is varied.}
\label{fig:nr-of-photons}
\end{center}
\end{figure}

\begin{table}[ht]
\caption{The table shows more precise values for the number of scattered photons in Fig.\ \ref{fig:nr-of-photons}, for some different laser powers. $N_{\gamma\gamma}$ is the number of generated QED scattered photons. We see that for a system like ELI a very large fraction of photons will be generated through the nonlinear quantum vacuum.} 
%\begin{tabular}
		\begin{tabular}{crr}
		\hline\noalign{\smallskip}			
		       Source beam power [PW] & $N_{\gamma\gamma}$ \\[1mm]
			\noalign{\smallskip}\hline\noalign{\smallskip}
			1 & 0.066 \\
			5 & 8.3 \\
			10 & 66 \\
			20 & $5.3\times 10^{2}$ \\
			50 & $8.3\times 10^{3}$ \\
			100 & $6.6\times 10^{4}$ \\
			1000 & $6.6\times 10^{7}$ \\
	       \noalign{\smallskip}\hline
		\end{tabular}
%\end{tabular}
\label{tab:nr-of-photons}
\end{table}
There are more recent proposals for detection of photon--photon scattering using four-wave interactions. \cite{Lundstrom-etal,Lundin-etal} has done more detailed calculations concerning experimental constraints, showing the feasibility of such an experiment. Using a proposed setup according to Fig.\ \ref{fig:3d}, the number of photons generated through a four-wave mixing process was determined. For simplicity, in the 3D case the interaction region will be modeled by a cube with side $b$, existing during a time $L/c$ (more precise numerical estimates can be derived using more accurate laser pulse profiles, but the result does not differ significantly from the one below). The estimated number of scattered photons per shot is found to be (see Fig. \ref{fig:nr-of-photons})
%\begin{widetext}
\begin{equation}\label{eq:number}
	%N_{3d}=1.31\eta^{2} G_{3d}^{2}\left(\frac{1\,{\rm \mu m}}{\lambda_{4}}\right)^{3}\left(\frac{L}{1\,{\rm\mu m}}\right)\left(\frac{P_1P_2P_3}{1\,{\rm PW}^3}\right)
	N_{\gamma\gamma}=\frac{2^7\pi\kappa^2}{\hbar c^4}\eta^2G_{3D}^2L\lambda_4^{-3} P_1 P_2 P_3,
	%\left(\frac{P_{2}}{1\,{\rm PW}}\right)\left(\frac{P_{3}}{\,1{\rm PW}  }\right),
\end{equation}
%\end{widetext}
where $\kappa =2\alpha ^{2}\hbar ^{3}/45m_e^{4}c^{5}$ defines the strength of the non-linear coupling in QED, $P_{j}$ is the power of the incoming pulses, $\lambda_{4}$ the generated wavelength determined by the resonance conditions and the pump laser parameters, $G_{3d}$ is a geometric factor capturing the polarization dependence and $\eta^{2}$ depends on the pulse model. Expressions for $G_{3d}$ and $\eta^{2}$ can be found in \cite{Lundin-etal}. Here we assume the same focal spot size independent of beam power. For further discussion on noise sources and their treatment, see Ref.\ \cite{Lundin-etal}.  

Experiments along the same lines as described for four-wave mixing above can also be used for a large number of other, non-QED tests, such as axion search \cite{Bernard,Dupays-etal,Bradley-etal}. Thus, progress of low-energy QED experiments could also prove to be useful for, e.g. dark matter searches.

\subsection{Is it worthwhile to further test QED?}
Would it be worthwhile to experimentally test the process of elastic photon-photon scattering? Elastic photon-photon scattering is a fundamental untested process in QED. An experiment is thus of great fundamental interest. This process can only become experimentally accessible using high power light sources. The principle of an experimental scheme is clean and simple; collide three beams satisfying a given matching condition in vacuum and a fourth beam will be generated. There are still practical problems that need to be considered, but an experiment of this kind could serve as a benchmark for more advanced (QED) high intensity experiments.

A high precision photon-photon scattering experiment may also offer the prospect that new physics could be probed. For instance, it may be possible to trace a deviation from the predicted polarization dependence to a deviation from Lorentz invariance. Lorentz invariance has not been tested in laboratory, although limits of any possible deviation has been presented based on astrophysical observations \cite{Kostelecky2001,Kostelecky2002}. In general one speaks of Born-Infeld type Lagrangians when we have a Lagrangian built from the Lorentz field invariants, and the coefficients are kept arbitrary \cite{born-infeld,dirac}. A Born-Infeld type Lagrangian may be written as
\begin{equation}\label{eq:born-infeld}
    {\cal L} = \xi^{2}\left[1-\sqrt{1+\frac{1}{2\xi^{2}}\left(E^{2}\!-B^{2}\right)-\frac{1}{16\xi^{4}}\left(\textbf{E}\cdot\textbf{B}\right)^{2}}\right],
\end{equation}
and this can give us some insight in how to interpret a deviation. Here $\xi$ is the relevant coupling constant for the modified electrodynamics. The Born-Infeld form of Lagrangian (\ref{eq:born-infeld}) also occurs as the effective field limit of quantized strings \cite{fradkin-tsetylin} (see Ref.\ \cite{tsetylin} for a review). In such effective Lagrangians the field strength coefficients contain the string tension. A Lagrangian of this form affects the geometric factor $G_{3d}$, and thus the polarization dependence of the signal seen in Fig.\ (\ref{fig:nr-of-photons}b).

Another feature of QED that has not been very well investigated is derivative correction to the Heisenberg Lagrangian due to field variations \cite{Mamaev}. The Lagrangian correction is a dispersive correction that can be written as
\begin{eqnarray}\label{eq:LD}
	\mathcal{L}_{D}=\sigma \varepsilon _{0}\left[F_{ab}\Box
F^{ab} -\left(\partial _{a}F^{ab}\right) \left( \partial _{c}F_{\phantom{b} b}^{c}\right)\right],
\end{eqnarray}
where $\Box =\partial _{a}\partial ^{a}=\left(c^{-2}\partial^2_t-\nabla^2\right)$ is the d'Alembertian and $\sigma =(2/45)\alpha c^{2}/\omega _{e}^{2}$ is the coefficient of the derivative correction. A slightly different and more general Lagrangian derivative correction than (\ref{eq:LD}) can be found in Refs.\ \cite{Gusynin1999,Gusynin1996}. In the presence of a strong magnetic field in vacuum, the vacuum will not only be birefringent but also dispersive. Ref.\ \cite{Lundin2007} also shows that the magnitude of the dispersive effects will be different for the two photon modes.

\section{Pair production}
Electron positron pair creation in vacuum is interesting to study because it is truly a non-perturbative effect of QED. There are two basic mechanisms behind this process. On one hand, we have the Schwinger mechanism which allows pairs to be produced if a static external electric field exceeds the critical field strength $E_{\text{crit}}\approx 10^{16}\ \mathrm{V/cm}$. On the other hand, we may have dynamical pair creation if the photon frequency is larger than the Compton frequency. Thus, the Schwinger threshold for pair creation needs to be modified for external fields with spatial and temporal variation. Both processes are exponentially suppressed in weak field regimes.

Since the field of a laser pulse is not static and uniform but slowly varying compared to the Compton frequency and the Compton wavelength, both the Schwinger and the dynamical pair creation processes are contributing to an actual threshold. The combined contribution from a strong slowly varying field and weak but fast electromagnetic wiggles has been studied recently in \cite{Schutzhold2008}. In general, spatial compression of a pulse tends to give a higher threshold for pair creation \cite{Narozhnyi, Gies2005}, whereas compression in time tends to lower the threshold. Several studies of time-varying fields have been performed suggesting that pair creation could in principle be detected in experiments with field strengths 1-2 orders of magnitude lower than the Sauter-Schwinger limit \cite{Alkofer2001,Narozhny2004,Ringwald2001,Ringwald2003,Hebenstreich2008}. Some authors \cite{Blaschke2006} have even suggested a yet lower threshold. It is, however, unclear to what extent spatial compression of a focused laser pulse will counter the positive effect of time-compression. Thus, the true threshold for detection of pair production in a laser experiment is still unknown. A new approach to resolve such non-perturbative problems is needed, and the semi-classical approach of world line instantons \cite{Dunne2005, Dunne2006} may offer means to do so. However, it is possible that successful experiments will come before a useful theory is at hand.

Pair creation experiments may also offer connection with high-energy physics. $e^+e^-\rightarrow \mu^+\mu^-$ is one of the fundamental processes in high-energy physics, and Ref.\ \cite{Muller2006} studies $\mu^+\mu^-$ creation by the interaction of a strong laser field with a low energy $e^+e^-$ plasma.

%Tanke, rumslig komprimernig av pulsen -> högre tröskelvärden för kollaps. Kollaps ger mkt komprimerade pulser -> undertrycker det parbildning? This has not been investigated.

\section{Pulse collapse}
Self interaction of a plane wave propagating in vacuum is prohibited in QED since the Lorentz field invariants vanish. However, light-light interactions may have a self focusing effect if the beam and/or background is properly modulated. For instance, self focusing can lead to collapse for two counter propagating beams, if the condition
\begin{eqnarray}
	\left(\frac{\alpha}{90\pi}\right)^{1/2}\frac{\left|E\right|}{E_{\text{crit}}}>\frac{r_p}{\lambda_p},
\end{eqnarray}
is satisfied. Here $r_p$ is the pulse radius and $\lambda_p = 2\pi/k_p$ the pulse wavelength. We may also have collapse for waves propagating caught in a wave guide \cite{Brodin2003,Shukla-OptComm}, or in a photon gas \cite{Marklund2003}, provided that the background and/or pulse is intense enough. A governing set of equations similar to that given by a Zakharov system \cite{Zakharov,Malomed}, where the radiation pressure of the pulse excites acoustic disturbances $\delta\mathcal{E}$ in the photon gas of density $\mathcal{E}_0$ which leads to a back reaction on the pulse with amplitude $E$ and wavenumber $k_p$ and subsequent collapse \cite{Marklund2003}, can be derived (here for the polarization averaged case)
\begin{subequations}
\begin{eqnarray}
&&	i\frac{\partial E}{\partial t} + \frac{c}{2k_p}\nabla_{\bot}^2E + \frac{11\alpha ck_p}{135 \pi} \frac{\mathcal E}{\epsilon_0 E_\mathrm{crit}^2} E = 0,  \\[2mm]
&&	\left(\frac{\partial^2}{\partial t^2} - \frac{c^2}{3}\nabla^2\right)\delta\mathcal{E} = -\frac{11\alpha\mathcal{E}_0}{45 \pi} \frac{c^2}{3} \nabla^2\left( \frac{\left|E\right|}{E_\mathrm{crit}} \right)^2.
\end{eqnarray}
\end{subequations}
These results are obtained using the one-loop approximation, but it is possible to also include higher order loops in the analysis \cite{Kharzeev2007}.

In a pulse collapse, field intensities and gradients can grow very large, and consequently higher order QED and derivative \cite{Mamaev} effects will become important for the dynamics. It is expected that pair creation will at some point attenuate the pulse intensity. The threshold for pair creation 
in such a spatially compressed pulse is, however, unresolved.

\section{Harmonic generation}
A virtual particle pair may absorb several laser photons, while producing a smaller number of photons as they annihilate each other. Consequently, the interaction of a laser pulse with quantum vacuum fluctuations can lead to generation of higher harmonics. Ref.\ \cite{DiPiazza2005} has studied generation of higher harmonics in a standing wave laser field and obtained the harmonic spectrum for different photon energies. Ref.\ \cite{Fedotov2006} on the other hand considers generation of higher harmonics from a focused intense laser pulse, and shows that this could in principle be detected already at field intensities of the order of $10^{27}\ \textrm{Wcm}^{-2}$. This is interesting since these harmonics will be generated at any facility producing ultra intense light pulses. %It is possible that at these high intensities non-perturbative effects may alter these results.

\section{Quantum plasmas}
\begin{figure}
  \includegraphics[width=0.9\columnwidth]{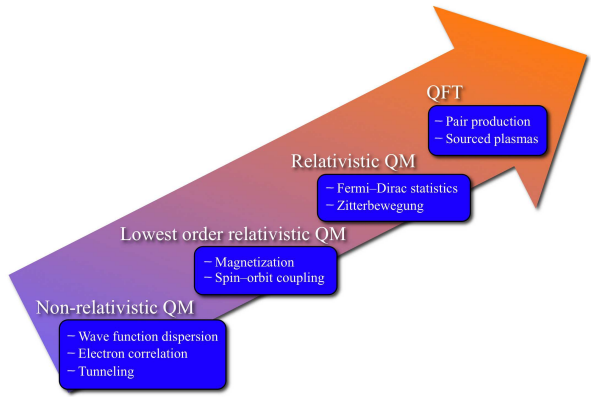}
  \caption{Starting from the simplest description of a quantum system in terms of a scalar Schr\"odinger equation, a continuous route connects this to quantum field theory, and thus gives an overall framework for dealing with collective quantum dynamics as well.}
  \label{fig:figure8}
\end{figure}

The field of quantum plasmas is a rapidly growing field of research.
From the non-relativistic domain, with its basic description in terms 
of the Schr\"odinger equation, to the strongly relativistic regime, with its
natural connection to quantum field theory, quantum plasma physics
provides promises of highly interesting and important application,
fundamental connections between different areas of science, as well
as difficult challenges from a computational perspective (see the mind map in Fig. \ref{fig:figure8}). The necessity to 
thoroughly understand such plasmas motivates a reductive principle of
research, for which we successively build more complex models based
on previous results. The simplest lower order effect due to 
relativistic quantum mechanics is the introduction of spin, and as such thus provides a first step towards a partial description of relativistic quantum plasmas. 

Already in the 1960's, Pines studied the excitation spectrum of quantum
plasmas \cite{Pines,pines-book}, for which we have a high density and a low temperature
as compared to normal plasmas. The theory for such plasmas can be viewed as a natural generalization to the theory of quantum liquids \cite{Pines-Quantum_liquids,Legget}. In such plasma systems, the finite width of the
electron wave function, among other things, makes quantum tunnelling effects crucial, leading to e.g.\
an altered dispersion relation. Since the pioneering work by Pines et al., a
number of theoretical studies of quantum statistical properties of plasmas
has been done (see, \textit{e.g.}, Ref.\ \cite{kremp-etal} and references therein). For
example, Bezzerides \& DuBois presented a kinetic theory for the quantum
electrodynamical properties of nonthermal plasmas \cite{bezzerides-dubois},
while Hakim \& Heyvaerts presented a covariant Wigner function approach for
relativistic quantum plasmas \cite{hakim-heyvaerts}. Recently there has been
an increased interest in the properties of quantum plasmas \cite
{Manfredi2005,haas-etal1,haas,shukla,garcia-etal,collection2C,collection2G,collection2H,collection2I,collection2K,collection2L,Haas-HarrisSheet,marklund-brodin,brodin-marklund,BM-pairplasma,%
shukla-eliasson2,shaikh-shukla,brodin-marklund2}%
. The studies have been motivated by the development in nanostructured
materials \cite{craighead} and quantum wells \cite{manfredi-hervieux}, the
discovery of ultracold plasmas \cite{li-etal} (see Ref.\ \cite{fletcher-etal}
for an experimental demonstration of quantum plasma oscillations in Rydberg
systems), astrophysical applications \cite{harding-lai}, or a general
theoretical interest. Moreover, it has recently been experimentally shown
that quantum dispersive effects are important in inertial confinement
plasmas \cite{glenzer-etal}. The list of quantum mechanical effects that can
be included in a fluid picture includes the dispersive particle properties
accounted for by the Bohm potential \cite
{Manfredi2005,haas-etal1,haas,shukla,garcia-etal,collection2C,collection2G,collection2H,collection2I,collection2K,collection2L,Haas-HarrisSheet}, the zero temperature
Fermi pressure \cite{Manfredi2005,haas-etal1,haas,shukla,garcia-etal}, spin
properties \cite{marklund-brodin,brodin-marklund,BM-pairplasma},
certain quantum electrodynamical effects \cite{Marklund-Shukla,Lundin2007,Lundstrom-etal,Brodin-etal-2007}, nano-plasmonics devices \cite{Marklund-EPL}, as well as quantum effects in the classical regime \cite{brodin-marklund-manfredi,brodin-prl08}. 
Within such descriptions, \cite
{Manfredi2005,haas-etal1,haas,shukla,garcia-etal,marklund-brodin,brodin-marklund,Lundin2007,Lundstrom-etal,Brodin-etal-2007} a unified picture of quantum and classical collective effects can be obtained.

\ \\

\section{Conclusions}

In this paper, we have mentioned but a few of the possible investigations of the quantum vacuum through the use of high-power lasers. In particular, focus has been given certain experiments that are feasible with current techniques, such as probing elastic photon--photon scattering. However, many important aspects of quantum dynamics and quantum vacuum physics have been left out of the discussion. In particular, effects such as 

\begin{itemize}

\item gigagauss laboratory magnetic field generation through solid target-laser interactions (see e.g.\ \cite{Mourou1998}),

\item photon splitting \cite{Adler1971}, 

\item Landau quantization \cite{Hanneke-etal} and its collective equivalent, and

\item the Unruh effect.

\end{itemize}
Here, perhaps the last point has attracted the largest interest (see e.g.\ \cite{unruh,chen:1999,Schutzhold,Brodin-cqg} and references therein) due to its heuristic coupling to the famous Hawking radiation \cite{hawking} and dynamical Casimir effect (see e.g.\ \cite{Mendonca2008} for a discussion). Due to its nature, the semi-classical concept of Hawking radiation ties together spacetime structure, thermodynamics, and quantum field theory into a surprising and fundamental bundle of ideas. A proper test of the Unruh effect could indeed constitute a breakthrough in our understanding of the deep connection between different areas of physics.

\acknowledgments

This work is supported by the European Research Council under Constract No.\ 204059-QPQV, and the Swedish Research Council under Contract No.\ 2007-4422. The authors  acknowledge the support by the European Commission under contract ELI PP 212105 in the framework of the program FP7 Infrastructures-2007-1.

\end{document}